\documentstyle[prb,aps,multicol,epsfig,amsmath,amsfonts,psfrag,rotating,pst-feyn,pst-key]{revtex}
   
 \newcommand{\nts}[1]{\tmspace{-}{#1\thinmuskip}{#1\txtmu}}
 \newcommand\eqlabel[1]{\label{#1}}

\allowdisplaybreaks
\begin{document}
\draft

\title{Determination of the Chemical Potential and the Energy \\
  of the \boldmath$ \nu=1/2 \; $\unboldmath FQHE System for low Temperatures} 

\author{J. Dietel}
\address{
Institut f\"ur Theoretische Physik, Universit\"at Leipzig,\\
Augustusplatz 10, D 04109 Leipzig, Germany }
\date{\today}
\maketitle
\begin{abstract}
We consider the energy density of a spin polarized $ \nu=1/2 $
system for low temperatures. We show that due to the elimination of
the magnetic field and the field of the positive background charge
in the calculation of the grand canonical potential of Chern-Simons systems 
through a
mean field formalism one gets corrections
to the well known equations which determine the chemical potential and the
energy from the grand canonical potential. We use these
corrected equations to calculate
the chemical potential and the energy of the $\nu=1/2 $ system at
low temperatures in two different approximations.  
\end{abstract}


\begin{multicols}{2}
\narrowtext

\section{Introduction}
The combination of an electronic interaction and a strong magnetic 
field in a two-dimensional electron system yields a rich variety of 
phases. These are best classified by the filling factor $ \nu $, 
which is the electron density divided by the density of a completely 
filled Landau level.
In this work we mainly consider energy calculations on   
systems with filling factor $\nu=1/2$.
These system are most suitably described
by the Chern-Simons theory. 
Within this theory one gets new quasi-particles
(composite fermions). In the case of
filling fraction $ \nu=1/2 $ every electron gets two magnetic flux quantums
to build a composite fermion  which does not see any
magnetic field in first approximation (mean field).
A field theoretical language for this scenario was first established by
Halperin, Lee, Read (HLR) (1992) \cite{hlr} as well as Kalmeyer and Zhang
(1992) \cite{ka1} for the $ \nu=1/2 $ system. The interpretation of
many experiments supports
this composite fermion picture. We mention transport experiments with
quantum (anti-) dots \cite{kan1}, and focusing experiments \cite{sm1} here.
An overview of further experiments can be found in \cite{wil2}.\\
HLR studied many physical quantities within the
random-phase approximation (RPA). Most prominent among these is the effective
mass of the composite fermions which they found to diverge at the Fermi
surface \cite{hlr,st1}.
Besides the theory of HLR there are
other alternative formulations of the composite fermionic picture
which are mainly based on a gauge transformation of the Chern-Simons
Hamiltonian \cite{sh1,al1}.

In the following we will apply the Chern-Simons theory of HLR to
calculate the chemical potential $ \mu $ and the ground state 
energy $ U $  
of the spin
polarized $ \nu=1/2 $ system for low temperatures $ T \ge 0 $. 
In \cite{di1,di2} we calculated the 
grand canonical potential of the $ \nu=1/2 $ system 
for temperature $ T=0 $ in RPA. The energy was calculated
from this potential  
by fixing the chemical potential to the value of a free electron system. 
Since the RPA consists of no anomalous Feynman graphs 
this is motivated as a good approximation for $ T= 0 $  by the 
Luttinger-Ward theorem \cite{lu1}. For temperature $ T \ge 0 $ one has to 
determine the chemical potential $ \mu $ by other methods. 
In the thermodynamic theory the chemical potential $ \mu $ and the 
energy $ U $ is calculated by the equations 
$ -\partial/(\partial \mu)\,\Omega =\rho $ and 
$ \partial/(\partial \beta) (\beta \Omega) +\mu \rho = U $ from the grand
canonical potential $\Omega$. 
$ \rho $ is the density of the electrons. We calculated in
\cite{di1,di2} the grand canonical  
potential of the Hamiltonian
of electrons in a magnetic field $ B $ subjected by an electron-electron 
interaction  
and an interaction with a positive background field $ \rho_B $
as a function of $ \mu $, $ \beta $. 
We eliminated  $ B $  
and $ \rho_B $ from the grand canonical potential $ \Omega $ by 
the constraint 
$ B=2 \pi \tilde{\phi} \rho $ and $ \rho_B= \rho $. 
By doing this, it is not clear whether the relations
$ -\partial/(\partial \mu)\Omega =\rho $ and 
$\partial/(\partial \beta) (\beta \Omega) +\mu \rho = U $ are further 
valid for the potential where these two external variables are 
eliminated through an implicit function.
We notice that 
the  elimination of $ \rho_B $ by $\rho_B= \rho $
is also a standard method in calculating the grand canonical potential 
of the Coulomb gas \cite{ge1}. We will show in section II 
that the above equations for the determination of $ \mu $ and $ U $  
from the grand canonical potential is further valid in the case of the  
Coulomb gas. In the case of the
$ \nu=1/\tilde{\phi} $
Chern-Simons gas we will show that the above relations get additional terms
of correction which are not too small.
We will show that these additional terms can be calculated from the    
variable eliminated grand canonical potential.

In \cite{di1,di2} we calculate the $ e^2 $-part 
of the ground state energy where $ e^2 $ is the coupling constant of the
Coulomb potential $ V^{ee}=e^2/r $. This is the first term of an
$ e^2$-expansion of the Coulomb energy and agrees  with the
Coulomb energy calculated for electrons which 'live' in the
lowest Landau level. We now have the following two options to expand the
calculation to low temperatures $ T >0 $. 
\begin{eqnarray}
& &  e^2 \sqrt{\rho}   \,  \ll \, \frac{1}{\beta} \, \ll  \, \frac{\rho}{m} ,\eqlabel{10} \\ 
 & & \frac{1}{\beta}\,  \ll \, e^2 \sqrt{\rho} \, \ll \,  \frac{\rho}{m} \,. \eqlabel{15}
\end{eqnarray}
Here $ m $ is the band mass of the electrons and
$ \beta=1/(k_B T) $ where $ k_B $ is the Boltzmann constant.
The first expansion (\ref{10}) results in an expansion in $ m/(\rho^2\beta) $
for the temperature corrections of every $ T=0 $-term.
The second expansion (\ref{15}) results in an expansion in  $ 1/(e^2
\sqrt{\rho} \beta) $ for the temperature corrections.

In a perturbational treatment of the Chern-Simons theory
the grand canonical potential 
is given by  the first expansion (\ref{10})
(one calculates the grand canonical
potential around a free grand canonical potential $  \Omega_0 $. The low
temperature expansion of $  \Omega_0 $ correspondence to an expansion which is
valid in the parameter range (\ref{10})).
Thus we will calculate the grand canonical potential at first
in the Hartree-Fock approximation.
It is well known \cite{ho1}, that the chemical potential
(for $ \tilde{\phi}=2$)
respectively the ground state energy contains exponentially 
vanishing temperature
corrections for small $ T $ for a system of electrons in a magnetic field
without Coulomb interaction. One could not obtain this feature by
perturbational methods. This behaviour of the temperature corrections 
of the chemical potential and the energy is no longer valid for electrons
in a magnetic field taking into consideration  the Coulomb interaction.
Therefore, in the following we will calculate only the Coulomb part
of the ground state energy. 
It is an interesting property of the Coulomb energy of the
Chern-Simons gas that the Coulomb exchange graph yields most of the energy
within the RPA \cite{di2}.
We will calculate in this paper the Coulomb energy of the
Hartree-Fock approximation of the Chern-Simons theory for low
temperatures $ T>0 $. 
This graph was calculated by Isihara and Toyoda \cite{isi3}
for the case of the Coulomb gas including the spin degrees of freedom. 
There are two reasons for a difference between the Coulomb part
of the Hartree-Fock energy of the Chern-Simons gas and the Coulomb gas.
First, the Chern-Simons gas is spin polarized which is not
the case for the Coulomb gas. This results 
in a much less steeper slope of the Coulomb energy as a function
of the temperature for the Chern-Simons gas in comparison to the
Coulomb gas. Second, as mentioned above, we have
different equations
for the determination of the chemical potential and the energy for these two
systems. 
We will show that this difference decreases the slope of the energy curve
of the
Chern-Simons gas further in comparison to the Coulomb gas.
This decreasing of the slope can be explained by perturbational arguments.

It is possible to calculate the effective mass of an interacting  system
through comparing its specific heat with the specific heat of an electron gas
without Coulomb interaction for $ B=0 $.
By comparing the two  expansions (\ref{10}) and (\ref{15})
with the scaling of the effective mass
\cite{hlr} $ m^* \sim \sqrt{\rho}/e^2 $ of the Chern-Simons theory
we obtain that only the
second expansion
leads to the correct scaling.
HLR \cite{hlr} and Kim, Lee \cite{ki2} calculate
the effective mass in the perturbational Chern-Simons theory by
a resummation of the grand canonical potential through the RPA.
With the help of this grand canonical potential they got a
logarithmic diverging effective mass.
We will calculate the chemical potential and the energy in the
parameter range (\ref{15}) by using the temperature corrections
of the grand canonical potential calculated by 
HLR and Kim, Lee. We will get the same energy
as in the simplification \cite{ki2}
by setting the chemical potential on the
lowest Landau level and neglecting the additional terms of correction
in determining the energy from the grand canonical potential. \\

The paper is organized as follows:
In section II we calculate the equations determining the
chemical potential and the ground state energy from a grand canonical
potential of the $ \nu=1/\tilde{\phi} $ system where the fields $ B $ and
$ \rho_B $ are eliminated. 
In section III we calculate the Coulomb energy and the chemical potential
of the $ \nu=1/2 $ system for
temperatures $ T>0 $ within the two parameter ranges (\ref{10}) and (\ref{15}).

\section{The determination of the chemical potential and the energy in
  \boldmath$ \nu=1/\tilde{\phi}\;$\unboldmath Chern-Simons systems}
 
In this section we show that 
one gets new equations to 
determine the chemical potential and the energy from the grand
canonical potential in Chern-Simons theories where the external fields,
i.e. the magnetic field $ B $ and the density of the positive background $
\rho_B $ 
are eliminated by some mean field conditions. 
We will show further that this is not the case for the Coulomb gas. 
In the following 
we will consider interacting spin polarized electrons moving in two dimensions in a
strong magnetic field $ B $ directed in the positive $ z $-direction
of the system.
The electronic density of the system is chosen such that the lowest Landau
level of a non-interacting system is filled to a fraction
$ \nu=1/ \tilde{\phi}$ where $ \tilde{\phi} $ is an even number. We are mainly
interested in $ \tilde{\phi}=2 $. The Hamiltonian of electrons in a magnetic
field is given by 
\begin{eqnarray}
& & H(\vec{A},e^2)=\int d^2r
\bigg[\frac{1}{2m}\big|\big(-i\vec{\nabla}+\vec{A}\big
)\Psi(\vec{r})\big|^2 \eqlabel{1040}   \\
& &   +\frac{1}{2}  \nts{1} \int  \nts{2}d^2r'  \nts{1} \Big\{ 
\nts{2}: \nts{1}(|\Psi(\vec{r})|^2-\rho_B)
 V^{ee}(|\vec{r}-\vec{r}\,'|)(|\Psi(\vec{r}\,')|^2-\rho_B)
 \nts{2} : \nts{2} \Big\}\bigg].
 \nonumber 
\end{eqnarray}
Here $ \Psi^+(\vec{r}) $ creates (and $ \Psi(\vec{r}) $ annihilates) an
electron 
with coordinate $ \vec{r} $. $ : O : $ is the normal ordering
of the operator $ O $.
$ V^{ee}(r)=e^2/r $ is the Coulomb interaction
where $ e^2=q_e^2 /\epsilon  $. $ q_e $ is the charge of the electrons and
$ \epsilon $ is the dielectric constant of the background field $ \rho_B $.
$ \vec{A}(\vec{r}) $ is the vector potential \
$ \vec{A}=1/2 \,\vec{B} \times \vec{r} $ and $\vec{B} $ is a 
homogeneous magnetic field in z-direction $ \vec{B}=B \vec{e}_z $
where $ \vec{e}_z $ is the
unit vector in $z $-direction. 
We suppose throughout this paper that $ B $ is a positive number. 
We used 
the convention $ \hbar=1 $ and $ c=1$ in the above formula
(\ref{1040}). Furthermore, we set $ q_e=1 $ for the coupling of the magnetic
potential to the electrons.
After performing a Chern-Simons
transformation \cite{zh2} of the electronic wave function one gets the
Hamiltonian of the composite fermions as:
\begin{eqnarray}
& & H_{CS}(\vec{A},e^2)=\int d^2r
\bigg[\frac{1}{2m}\big|\big(-i\vec{\nabla}+\vec{A}  
+ \vec{a}_{CS}\big)\Psi(\vec{r})\big|^2 \eqlabel{1050}   \\
& &  +\frac{1}{2}  \nts{1} \int  \nts{2}d^2r'  \nts{1} \Big\{ 
\nts{2}: \nts{1}(|\Psi(\vec{r})|^2-\rho_B)
 V^{ee}(|\vec{r}-\vec{r}\,'|)(|\Psi(\vec{r}\,')|^2-\rho_B)
 \nts{2} : \nts{2} \Big\}\bigg].
 \nonumber  
\end{eqnarray}
The Chern-Simons operator $ \vec{a}_{CS} $ is defined
by $ \vec{a}_{CS}(\vec{r})=\tilde{\phi} \int d^2r'\vec{f}(\vec{r}-\vec{r}\,')
\Psi^+(\vec{r}\,')\Psi(\vec{r}\,') $. 
Here $ \Psi^+(\vec{r}) $ creates (and $ \Psi(\vec{r}) $ annihilates) a
composite fermion
with coordinate $ \vec{r} $.
The function $ \vec{f}(\vec{r}) $ is given by 
$ \vec{f}(\vec{r})=-\vec{e}_z \times \vec{r}/r^2 $.
From this Hamiltonian we obtain in the case of filling fraction
$ \nu=1/\tilde{\phi} $ and electro neutrality,
i.e. $ B=2 \pi \tilde{\phi} \langle \hat{\rho} \rangle $ and $ \rho_B=\langle \hat{\rho} \rangle $, that in the
perturbation theory of $ H_{CS} $ the Hartree couplings
$ \vec{A}+\tilde{\phi} \int d^2r' \,\vec{f}(\vec{r}-\vec{r}\,')
\langle\Psi^+(\vec{r}\,')\Psi(\vec{r}\,')\rangle $ and
$ \int
d^2r'\,V^{ee}(|\vec{r}-\vec{r}\,'|)
(\langle\Psi^+(\vec{r}\,')\Psi(\vec{r}\,')\rangle-\rho_B)
$ are zero. If we calculate the Feynman graphs of the theory under these
conditions we eliminate the external fields $ B $ and $ \rho_B $ 
of the partition function
$ Z=\mbox{Tr}[e^{-\beta (H_{CS}(\vec{A},e^2)-\mu \hat{N})}]=
\mbox{Tr}[e^{-\beta (H(\vec{A},e^2)-\mu \hat{N}}]$. 
Here
$ \mbox{Tr}[\cdot] $ is the trace of the argument.  
$ \hat{N} $ is the particle number operator
$\hat{N}=\int\,d^2r\,\Psi^+(\vec{r})\Psi(\vec{r}) $. 
The variables left in $ \Omega $ are $ \mu $ and $ \beta$.
In the following we will deal with three different 
systems with a growing degree of complication.
\\[0.1cm]

At first we deal with the Coulomb gas ($ \vec{A}=0 $ in (\ref{1040})). 
In perturbational calculations of this system
one often calculates the Feynman graphs
under the condition $ \rho_B=\langle \hat{\rho} \rangle $ 
\cite{ge1}.
$ \hat{\rho}(\vec{r}) $ is the density operator
$\Psi^+(\vec{r}) \Psi(\vec{r})$.
$ \langle\cdot \rangle $ is the expectation value of the Gibb's potential of
the Hamiltonian. For a homogeneous system we have 
$ \langle\hat{\rho}(\vec{r})\rangle =
\langle\Psi^+(\vec{r}) \Psi(\vec{r})\rangle=\langle\hat{\rho}\rangle$.
Thus one calculates a partition function $ Z'(\mu,\beta) $
from $ Z(\mu,\beta,\rho_B)=\mbox{Tr}[e^{-\beta(H(0,e^2)-\mu \hat{N})}] $ by  
\begin{eqnarray}
Z'(\mu,\beta)& := & Z\left(\mu,\beta,\rho_B(\mu,\beta)\right)
\;,\eqlabel{1200}\\
\rho_B(\mu,\beta) & := & \frac{1}{A\, \beta} \frac{\partial}{\partial \mu} 
\log \left(Z\right)(\mu,\beta,\rho_B(\mu,\beta))\;.\eqlabel{1210}
\end{eqnarray}
$ A$ is the area of the system. 
We employed in  (\ref{1200}) and  (\ref{1210}) the mathematical notation for the 
ordering of the derivation and insertion of the arguments of the functions. 
This
means for example for equation (\ref{1210}) that we have to partially 
derivate at first the
function Z depending on the variables $ (\mu,\beta,\rho_B)$. 
Afterwards 
we have to insert the expressions given in the function brackets. 
We have to point out explicitly  that
$ \rho_B(\mu,\beta) $ is a function of $ (\mu,\beta) $ which is 
defined by equation (\ref{1210}). We now want to determine the chemical
potential $ \mu $ and the energy $ U $ from $ Z' $. At first we deal 
with the derivation of $ Z $ with respect to $ \mu $: 
\begin{eqnarray}
\frac{1}{\beta}\frac{\partial}{\partial \mu} 
\log\left(Z\right)(\mu,\beta,\rho_B(\mu,\beta)) & = &  
 \frac{1}{\beta}\frac{\partial}{\partial \mu} 
 \log\left(Z'\right)(\mu,\beta)+K_1 \nonumber \\
 & = &  A \,\langle\hat{\rho}\rangle\;.    \eqlabel{1230}
\end{eqnarray}
$ K_1 $ is defined by  
\begin{eqnarray}
K_1 & = & -\int\limits_A \int\limits_A d^2r d^2r' 
(\langle\hat{\rho}(\vec{r})\rangle-\rho_B)
  V^{ee}(\vec {r}-\vec{r}\,')    \nonumber \\ 
& & \left. \times 
\frac{\partial}
{\partial \mu} 
\rho_B(\mu,\beta)\right|_{\langle\hat{\rho}\rangle=\rho_B} =0\;.
\eqlabel{1220}
\end{eqnarray}
Thus we get the following equations determining the chemical
potential $ \mu $ and the energy $U $ of the Coulomb system   
$ -\frac{\partial}{\partial \mu} \Omega'(\mu,\beta)
 \; = \; \langle \hat{\rho}\rangle  $ and $ 
U=\frac{\partial}{\partial \beta} (\beta \Omega')(\mu,\beta)+
\mu \langle \hat{\rho} \rangle $. The grand
canonical potential $ \Omega'(\mu,\beta) $ is defined by
$ \Omega'=-1/(\beta A) \log(Z') $. From these equations
we obtain  that in the case of the Coulomb gas we get no correction to the
well known equations for the determination of the chemical potential and the
energy. \\[0.1cm]

In the following we will derive equations to  
determine  $ \mu $ and $ U
$  for  
the $ \nu=1/\tilde{\phi} $ Chern-Simons gas
with no Coulomb interaction and $ T=0 $ . 
As mentioned earlier the partition function
$ Z(\mu,\beta,B)=\mbox{Tr}[e^{-\beta(H (\vec{A},0)-\mu \hat{N})}]  $ of this system 
is usually 
calculated under the constraint
$ \langle\hat{\rho} \rangle = B/(2 \pi \tilde{\phi}) $ giving $ Z'(\mu,\beta)$. So $ Z' $ is defined by 
\begin{eqnarray}
Z'(\mu,\beta)&:= & Z\left(\mu,\beta,B(\mu,\beta)\right)\, , \eqlabel{1270}\\
B(\mu,\beta)& := &  
(2\pi \tilde{\phi})\frac{1}{A \, \beta} 
\frac{\partial}{\partial \mu} \log \left(Z\right)
\left(\mu,\beta,B(\mu, \beta)\right) \;. \eqlabel{1280}
\end{eqnarray}

\begin{figure}[t]
 { \psset{unit=0.3cm}
   \begin{center}
\begin{pspicture}(10,8)
\psset{linewidth=1.3pt,arrowinset=0}
\psframe(1,8.5)(5,6)
\psframe(1,4)(5,1.5)
\psline[ArrowInside=-](2,1.5)(2,0.5)
\psline[ArrowInside=-](4,1.5)(4,0.5)
\WavyLine[n=20,beta=180](3,4)(3,5.5)
\psline[ArrowInside=-](3,5.5)(3,6)
\rput(3,6.95){\footnotesize{$ \hat{\vec{j}}(\vec{r}) $ }}
\rput(3,3.3){\footnotesize{$ \hat{\rho}(\vec{r}\,') $ }}
\rput(6.0,5){\footnotesize{$ \tilde{\phi}\,\vec{f}(\vec{r}-\vec{r}\,') $ }}
\end{pspicture}
\end{center}
\vspace{-0.3cm}
\caption{Density graphs, which contribute to the $ \mu $-correction. \label{bildm}}
}
\end{figure}
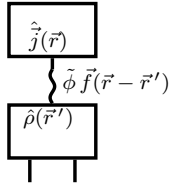

From these definitions we get for the derivate of the partition function
$ Z $ with respect to $ \mu $ 
\begin{eqnarray}
\frac{1}{\beta}\frac{\partial}{\partial \mu} \log(Z)(\mu,\beta,B(\mu,\beta))
&=&  \frac{1}{\beta}\frac{\partial}{\partial \mu}\log(Z')(\mu,\beta)+K_2
\nonumber \\
& = &  A \langle\hat{\rho} \rangle \;. \eqlabel{1300}   
\end{eqnarray}
$ K_2 $ is defined by 
\begin{equation}\eqlabel{1290}
K_2 =  - \frac{1}{2\pi}\int\limits_A \int\limits_A d^2r\, d^2r'\, 
\langle\hat{\vec{j}} \rangle(\vec{r}) \vec{f}(\vec{r}-\vec{r}\,')\frac{\partial}
{\partial \mu} B(\mu,\beta) \;. 
\end{equation}
$ \hat{\vec{j}}(\vec{r}) $ is the second quantized current operator 
$ (1/2m)( \Psi^+(\vec{r})(-i \vec{\nabla}+\vec{A}(\vec{r}))\Psi(\vec{r})
+[(-i \vec{\nabla}+\vec{A}(\vec{r}))\Psi(\vec{r})]^+\Psi(\vec{r})) $.
It is clear that the expectation value of the current operator in (\ref{1290})
is built with respect to the Gibb's operator of the Hamiltonian $ H(\vec{A},0)$
(\ref{1040}). 
At first glance one may think that $ K_2 $ is zero because we have 
$\langle\hat{\vec{j}} \rangle(\vec{r})=0 $ for $ A \to \infty $. We have to be
careful with such argumentation because the multiplicative term
$\vec{f}(\vec{r}-\vec{r}\,') $ 
in (\ref{1290}) is not integrable for $ |\vec{r}-\vec{r}\,'| \to \infty  $.
The right way to calculate $ K_2 $ is to evaluate 
$\langle\hat{\vec{j}} \rangle(\vec{r}) $ for a finite system. 
Then after integrating (\ref{1290}) for this finite system
one should take the limit $ A \to \infty $. 
For a finite system we have nothing but
a current at the edge of the system. This
ring current was calculated perturbationally by Shankar and Murthy \cite{sh1}. 
With the help of this ring current one could calculate $ K_2 $. In the
following we will calculate $ K_2 $ by another method which results in the
same value for $ K_2 $ but which does not use any perturbational results. 
Using the one particle eigen functions  $ u_{0l} $ of an electron in a
magnetic field in the lowest Landau
level (symmetric gauge) we get 
\begin{equation}\eqlabel{1330}
-\frac{1}{2\pi} 
\int\, d^2r\, d^2r' \left<u_{0k}|\hat{\vec{j}}(\vec{r})|u_{0l}\right>
\vec{f}(\vec{r}-\vec{r}\,')=\frac{1}{2m}\;\delta_{k,l} \;.
\end{equation}
In (\ref{1330}) we supposed that $ u_{0k}$ is orthogonal
to $ u_{0l}$ for $ k \not=l $. 
For deriving (\ref{1330}) we used 
$ \vec{A}(\vec{r})=-B/(2\pi)\int d^2r'\vec{f}(\vec{r}-\vec{r}\,')
=B/2 (-y,x) $.   \\
By using (\ref{1280}), (\ref{1300}), (\ref{1290}) and (\ref{1330}) 
we get  
\begin{equation}\eqlabel{1340}
-\frac{\partial}{\partial \mu}\Omega'(\mu,\beta) 
 +\frac{\pi \tilde{\phi}}{m}\,
 \langle\hat{\rho} \rangle\,\frac{\partial}{\partial \mu}
\langle \hat{\rho} \rangle(\mu,\beta)= \langle \hat{\rho} \rangle    \;.
\end{equation}
Here we used $ B(\mu,\beta)=2 \pi \tilde{\phi} \langle \hat{\rho} \rangle
(\mu,\beta)$.
So we have to solve a differential equation to get the chemical
potential for a given density $ \langle \hat{\rho} \rangle$.
For the  determination of the energy $ U $ from $ \Omega' $
we get with the help of a similar derivation as above 
\begin{equation}\eqlabel{1350}
U=\frac{\partial}{\partial \beta} (\beta \, \Omega')(\mu,\beta)
-\frac{\pi \tilde{\phi}}{m} \, \beta \langle \hat{\rho} \rangle 
\frac{\partial}{\partial \beta}\langle \hat{\rho} \rangle(\mu,\beta) +\mu \langle\hat{\rho} \rangle  \;.
\end{equation}
 Thus we obtain from the equation (\ref{1340}) and (\ref{1350}) 
 an additional
term in comparison to the equations which  determine the energy and
the chemical potential from the grand canonical potential
$ \Omega=-1/(\beta A) \log(Z) $.   
In figure \ref{bildm} we show density graphs of the Hamiltonian
$H_{CS}(\vec{A},e^2) $ which contributes to this additional term.
Since   $ \langle\hat{\rho} \rangle $ is finite for $ T=0 $ the 
second term in
(\ref{1350}) is zero. 
It is well known that the energy and the chemical potential
of electrons in a magnetic field at $ T=0 $ are given by   $ U=\mu \rho $ and
$ \mu=\pi\rho \tilde{\phi}/m $. 
So we get from (\ref{1340}) and (\ref{1350}) 
\begin{equation}\eqlabel{1370}
-\frac{\partial}{\partial \mu}\Omega'(\mu,\beta)\quad = \quad 0
\quad ,\quad 
\frac{\partial}{\partial \beta} (\beta\Omega')(\mu,\beta)\quad = \quad 0 \;. 
\end{equation}
We obtain from these equations that in the case of an interaction 
free Chern-Simons system 
the correction to the equations determining  $ \mu $ and $ U $ are not neglectable.
This should be also  the case for a Chern-Simons system taking into account
the 
Coulomb interaction. \\ [0.1cm]

In the following part of this section we will derive 
the equations for getting $ \mu $ and $ U$  of  
the $ \nu=1/\tilde{\phi} $ Chern-Simons gas for low temperatures
taking into consideration the  Coulomb interaction. 
In the mean field treatment of the $ \nu=1/\tilde{\phi} $ 
Chern-Simons theory one usually calculates
from the partition function $ Z(\mu,\beta,\rho_B,B)=\mbox{Tr}[e^{-\beta (H(\vec{A},e^2)-\mu
  \hat{N}) } ] $ 
the following reduced partition function 
\begin{equation} \eqlabel{1380}
Z'(\mu,\beta)=Z\left(\mu,\beta,\rho_B(\mu,\beta),B(\mu,\beta)\right) \;.
\end{equation}
$ \rho_B(\mu,\beta) $, $ B(\mu,\beta) $ are defined by 
(\ref{1210}), (\ref{1280}).
It is now possible to make the same derivations for this system as in the
case of the Chern-Simons system with no Coulomb interaction. By doing this we
get equation (\ref{1300}).
Unfortunately it is then hard to calculate the expectation value
(\ref{1290}) for this system because we have a mixing of
higher Landau levels.
Nevertheless one can get the equations which
determine  $ \mu $ and $ U $ by scaling arguments.
This will be done in the following.
From the definition of $ Z' $ we get 
\begin{eqnarray}
& & \frac{1}{A\,\beta}\frac{\partial}{\partial \mu} 
\log\left(Z\right)(\mu,\beta,\rho_B(\mu,\beta),B(\mu,\beta))
           \eqlabel{1390}      \\     
& & = \frac{1}{A\,\beta}\Bigg[
-\frac{\partial}{\partial B} 
\log\left(Z\right)(\mu,\beta,\rho_B(\mu,\beta),B(\mu,\beta)) \frac{\partial}{\partial \mu}
B(\mu,\beta)    \nonumber  \\
& & -\frac{\partial}{\partial \rho_B} 
\log\left(Z\right)(\mu,\beta,\rho_B,B(\mu,\beta)) \frac{\partial}{\partial \mu}
\rho_B(\mu,\beta)  \nonumber \\
& & +\frac{\partial}{\partial \mu} 
\log\left(Z'\right) (\mu,\beta)\Bigg] . \nonumber 
\end{eqnarray}
As in the case of the  
Coulomb gas the  second term on the right hand side is zero.
Thus we have to calculate the first term on the right hand side 
of equation (\ref{1390}). 
With the help of the definitions  
\begin{eqnarray}
& & \nts{10} a_{n p}=  \int d^2r\; u_{np}(\vec{r}) \Psi(\vec{r}) \;,\eqlabel{1392} \\
& &\nts{10} V^{ee}_{n_1 \, p_1,n_2 \, p_2, n_3 \, p_3, n_4 \, p_4}= \int d^2r d^2r'\;
V^{ee}(\vec{r}-\vec{r}\,')\nonumber \\  
& & \hspace{2cm} \times 
u^*_{n_1p_1}(\vec{r}) u_{n_2p_2}(\vec{r}) u^*_{n_3p_3}(\vec{r}\,')
u_{n_4p_4}(\vec{r}\,')  \;\nonumber  
\end{eqnarray}
the operators $ H_{0}(B) $ and $ H_{ee}(B) $ are defined by 
\begin{eqnarray}
H_{0}(B) & = & \sum\limits_{n,p} \, \frac{B}{m}\left(n+\frac{1}{2}\right) 
\, a^+_{n p} a_{n p} \;, \eqlabel{1395}\\
H_{ee}(B)&= & \frac{1}{2}\sum\limits_{n_1p_1,n_2 p_2,
       \atop n_3p_3,n_4p_4} 
       V^{ee}_{n_1p_1,n_2 p_2, n_3p_3, n_4p_4}  \eqlabel{1400}  \\
    & &   \times : \left(a^+_{n_1p_1}a_{n_2p_2}- \rho_B\right)
\left(a^+_{n_3p_3}a_{n_4p_4}- \rho_B\right): \;. \nonumber 
\end{eqnarray}
$u_{n p} $ are the one particle wave functions in the symmetric gauge for the
magnetic field $ B$ in the nth landau level  \cite{cha1}. 
With the help of these operators we get for the grand canonical potential
$ \Omega=-1/(\beta A) \log(Z) $ 
\begin{eqnarray}
& & \Omega(\mu,\beta,\rho_B,B)=-\frac{1}{\beta}\lim\limits_{A\to\infty} \frac{1}{A} \eqlabel{1410}  \\
& & \hspace{1cm} 
\times \log\left[\mbox{Tr}\left[e^{-\beta(H_{0}(B)+H_{ee}(B)
-\mu\,\hat{N})}\right]_{B,A}\right] \, .\nonumber 
\end{eqnarray}
$ \nts{1}\mbox{Tr}[...]_{B,A} $ is the trace of slater determinants  
of Landau functions which have their support in the
area $ A$. The number of Landau functions in this  area is 
proportional to $ B $ \cite{cha1}. 
With the help of a length scaling  $L \rightarrow L /\sqrt{B} $
(i.e. we implement  the trace in (\ref{1410}) in the Landau basis and make
the substitution $ \vec{r} \rightarrow \vec{r} /\sqrt{B} $) 
we get 
\begin{eqnarray}
& & \Omega(\mu,\beta,\rho_B,B)=-\frac{B}{\beta}\lim\limits_{A \to \infty}\frac{1}{A}\eqlabel{1420}  
 \\
& & \hspace{1cm} \times \log\left[\mbox{Tr} \left[e^{-\beta(B H_{0}(\frac{B}{|B|})
      +\sqrt{B}H_{ee}(\frac{B}{|B|})
-\mu\,\hat{N})}\right]_{\frac{B}{|B|},A}\right]    \,.\nonumber 
\end{eqnarray}
Thus we get 
\begin{eqnarray}
& &  \frac{\partial}{\partial B} \Omega(\mu,\beta,\rho_B,B) \eqlabel{1430}\\ 
& &\hspace{0.5 cm} =
\frac{1}{B}\left(\langle H_0(B) \rangle+\frac{1}{2} \langle H_{ee}(B) \rangle
  +\Omega(\mu,\beta,\rho_B,B)\right)  \,. \nonumber  
\end{eqnarray}
With the help of  (\ref{1390}) and $ \langle \hat{\rho} \rangle=1/(2 \pi \phi)
B(\mu,\beta) $ we get for the equation determining the chemical potential
$ \mu$ from $ \Omega'$ 
\begin{eqnarray}
& & \left(\langle H_0(B) \rangle+
  \frac{1}{2}\langle H_{ee}(B) \rangle
  +\Omega'(\mu,\beta)\right) \frac{1}{\langle\hat{\rho} \rangle} \frac{\partial }{\partial
  \mu}\langle\hat{\rho} \rangle(\mu,\beta)  \nonumber \\
& & \hspace*{0.1cm} -\frac{\partial}{\partial \mu} \Omega'(\mu,\beta)
= \langle\hat{\rho} \rangle  \;.
\eqlabel{1440}
\end{eqnarray}
Similar as above we get for the energy of the $ \nu=1/\tilde{\phi} $ system
\begin{eqnarray}
 U  \nts{1} &  = & \nts{1}-\Big(\nts{1}
  \langle H_0(B) \rangle \nts{0.5}+ \nts{0.5} \frac{1}{2}\langle H_{ee}(B) \rangle 
  \nts{0.5}+ \nts{0.5} \Omega'(\mu,\beta) \nts{1} \Big) 
  \beta \frac{1}{\langle \hat{\rho} \rangle }
\frac{\partial }{\partial \beta}\langle \hat{\rho} \rangle (\mu,\beta)
\nonumber \\
& & +\frac{\partial}{\partial \beta} (\beta\Omega')(\mu,\beta)
+\mu \langle \hat{\rho} \rangle  \;. \eqlabel{1450}  
\end{eqnarray}
In the case of a determination of the energy and the chemical potential to
order $ e^2 $ 
it is possible to get 
$ \langle H_0 \rangle+\frac{1}{2}\langle H_{ee} \rangle $ 
from $ \Omega' $ for low temperatures.
In the following we will derive this relation at first  for $ T=0 $. 
Afterwards we will generalize
the results to the case of low temperatures.
It is well known from  perturbation theory that  the ground state 
wave function
has only higher Landau level components scaling with $ e^2 $. So we get 
for temperature $ T=0 $ 
\begin{equation}\eqlabel{1460}
\langle H_0(B) \rangle =\pi \tilde{\phi} \frac{\langle\hat{\rho} \rangle^2}{m}+O(e^4)   \;.
\end{equation}
For $ T>0 $ it is no longer correct that the second  
term of the $ e^2 $-expansion of $ \langle H_0(B) \rangle $
scales like $ e^4 $. 
Nevertheless because of the discreteness of the Landau
 levels we obtain that these $ T $-corrections are exponentially vanishing for
 $ T \to 0 $.
The reason for this is that due to the expansions (\ref{10}) and (\ref{15})
there is a coupling factor $ e_0^2 $ such 
that we get no overlap between the energy eigen values of  wave functions
resulting from the lowest Landau level by Coulomb perturbations and  
wave functions resulting from higher Landau levels  for all $ e^2 \le  e_0^2
$.
Since $ 1/\beta \ll \rho/m $ for both expansions (\ref{10}) and (\ref{15})
we get that the kinetic energy resulting  from wave functions of higher
Landau levels 
by Coulomb perturbations are suppressed exponentially with
$ \exp[-\beta \rho/m] $. 
Since we are only interested in polynomial $ T $-contributions
 to the energy we can neglect these $ T $-corrections in (\ref{1460}). 
Under the consideration of equation (\ref{1460}) we can get 
$ \langle H_{ee} \rangle $ up to order $ e^2 $ from  $ \Omega $ by 
\begin{eqnarray}  
& &[ \langle H_{ee}(B) \rangle]_{e^2} = [U(\langle\hat{\rho} \rangle,\beta)]_{e^2}
\eqlabel{1480} \\
& & =  
\left[\frac{\partial (\beta\Omega)}{\partial \beta}
  (\mu(\langle\hat{\rho} \rangle,\beta),\beta,\langle\hat{\rho} \rangle,2\pi \tilde{\phi}\langle\hat{\rho} \rangle )
  +\mu(\langle\hat{\rho} \rangle,\beta) \langle\hat{\rho} \rangle\right]_{e^2}. 
\nonumber 
\end{eqnarray}
We denote by  $ [...]_{e^2} $  the $ e^2 $-part of the bracket for $ T=0 $.
In the case of low temperatures $ T $ we have to distinguish between the
two low temperature expansions (\ref{10}) and (\ref{15}). 
In the case of the temperature expansion (\ref{10})
$ [...]_{e^2} $ is given by the
$ e^2 (\rho)^{3/2}/(\beta \rho/m)^n $ ($n\ge0$)  terms of the bracket.  
For the temperature expansion (\ref{15}) 
$ [...]_{e^2} $ is given by the  $ e^2 (\rho)^{3/2}/(\beta e^2
\sqrt{\rho})^n $ ($n\ge0$) terms
of the bracket. We may  now determine the chemical potential and the energy
in every order of $ 1/\beta $ by using successive the equations 
(\ref{1440}), (\ref{1450}), (\ref{1460}), (\ref{1480}) under consideration
that
$ \lim_{\beta \to \infty}\frac{\partial}{\partial \beta}\langle\hat{\rho}\rangle
=O(1/\beta^{1+\epsilon}) $ for a number $ \epsilon $ with $ \epsilon >0 $.
Thus we can calculate the chemical potential and the energy
to every order in $ 1/ \beta $
through a finite number of successive insertions of these equations.  
Doing this for $ T=0 $ we get 
\begin{eqnarray}\eqlabel{1490}
& & \bigg(
  \frac{1}{2\langle\hat{\rho} \rangle}\left[
 \frac{\partial}{\partial\beta}(\beta\Omega')
 (\mu(\langle\hat{\rho} \rangle,\infty),\infty)
 +\mu(\langle\hat{\rho} \rangle,\infty) \langle\hat{\rho} \rangle\right]_{e^2}\nonumber\\
& &+\frac{1}{\langle\hat{\rho} \rangle} \Omega'(\mu,\infty)+(\pi \tilde{\phi}) \frac{\langle\hat{\rho} \rangle}{m}\bigg)
\frac{\partial}{\partial \mu} \langle\hat{\rho} \rangle(\mu,\beta) 
-\frac{\partial}{\partial
 \mu} \Omega'(\mu,\infty) \nonumber \\
& & 
\hspace{0.1cm} =\langle\hat{\rho} \rangle
\end{eqnarray}
and 
\begin{equation}\eqlabel{1500}
U=\frac{\partial (\beta\Omega')}{\partial \beta}(\mu,\infty) 
+\mu \langle\hat{\rho}  \rangle\;.
\end{equation}
By reducing equation (\ref{1490}) to the Chern-Simons gas without Coulomb
interaction we get a difference of a summand
$\frac{1}{\langle\hat{\rho}\rangle}\Omega'(\mu,\infty)\frac{\partial}{\partial\mu}
\langle \hat{\rho} \rangle (\mu,\infty ) $ between the equations (\ref{1340}) and
(\ref{1490}). The grand canonical potential for an electron gas
in a magnetic field without Coulomb interaction is given by 
 \begin{eqnarray}
& & \Omega_B(\mu,\beta,B)=-\frac{1}{2\pi}\frac{B}{\beta} \eqlabel{1510}\\
& & 
\times \sum\limits_{n} 
\log\left(1+\exp\left[-\beta\left(\left(n+\frac{1}{2}\right) 
\frac{B}{m}-\mu\right)\right]\right) \;. \nonumber 
\end{eqnarray}
The chemical potential at low temperatures  for
the $ \nu=1/2 $ gas is given by \cite{ho1} 
$ \mu=\frac{(2\pi \rho )}{m}+O(e^{-\beta \rho/m}) $.  
 Since $\lim_{\beta \to \infty}
\Omega'(\mu(\rho,\beta),\beta)= O(1/\beta)$  we obtain for $ T=0 $ that the
equations (\ref{1340}) and
(\ref{1490}) are in accordance.

\section{The chemical potential and the energy
  of the \boldmath $ \nu=1/2 $ \unboldmath Chern-Simons
  system for  \boldmath $ T \ge 0 $ \unboldmath}
In this section we will calculate the chemical potential and the
energy  of the $ \nu=1/2 $ system for low temperatures $ T $
and $ e^2 \sqrt{\rho} \ll \rho/m$. As written in the introduction
there are two parameter ranges  for an expansion of the energy and the
chemical potential. These are given in (\ref{10}) and (\ref{15}).
In the following we will calculate in subsection A the chemical potential and
the energy in the parameter range (\ref{10}). In subsection B
we will calculate the chemical potential and the energy
within the parameter range (\ref{15}).

\subsection{The chemical potential and the energy
  of the \boldmath $ \nu=1/2 $ \unboldmath Chern-Simons
  system for \boldmath $ e^2 \sqrt{\rho}   \,  \ll \, 1/\beta \, \ll  \,
 \rho/m $\unboldmath}

\begin{figure}
 \centerline{\psfrag{x}{\smash{\raisebox{-0.15cm}{\footnotesize $m k_B T\, r_s^2 $
       }}} 
\psfrag{y}{ \turnbox{180}
  {\small \hspace{-1.8cm} Energy Density $ U \, [e^2/r_s^3]$} }
  \epsfig{file=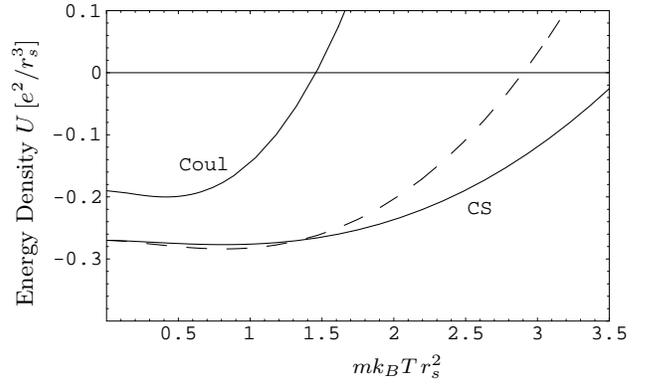,width=8cm}}
\hspace{0.2cm}
 \caption{The  $ e^2 $ part of the energy density
   of the Coulomb-
  (Coul) as well as the  $ \nu=1/2 $ Chern-Simons (CS) gas.   
 The energy density is given in units of  $e^2/r_s^3 $, where
 $r_s $ is the electron radius $ \rho=1/(\pi r_s^2) $.
 The curve Coul is the $ e^2 $-energy density of the
 two dimensional Coulomb gas (\ref{5650}) including the spin degree of freedom.  The curve CS is the $e^2 $-energy density of the  
 $ \nu=1/2 $ Chern-Simons gas (\ref{5640})
 originating
 from the spin polarized Coulomb exchange graph under consideration of
the equations (\ref{5610}) and (\ref{5620}). 
The dashed curve  is the $ e^2 $-energy density of a spin polarized
 two dimensional Coulomb gas. 
\label{energ}}
\end{figure}

As mentioned in the introduction the Chern-Simons perturbations theory
has its validity in the parameter range (\ref{10}). 
Thus it is possible to  calculate the energy and the chemical
potential in this parameter range within the Hartree-Fock
approximation. 
As mentioned in the last section one gets exponentially vanishing temperature
  corrections 
to  the chemical potential and the energy at low temperatures 
for  the $ \nu=1/2 $ 
system  without Coulomb interaction. They are not calculable by perturbational 
methods. Because of this we will calculate only the Coulomb 
part of the grand canonical potential perturbationally. The exact magnetic
part of the 
grand canonical potential is given by 
$ \Omega_B(\mu,\beta,(2 \pi \tilde{\phi} )\rho^*)$ 
(\ref{1510}).
In this expression $ \rho^* $ is defined by
$ \rho^*:= \partial/(\partial \mu)
\Omega_B(\mu,\beta,(2 \pi \tilde{\phi} )\rho^*)$. As we mentioned
above in the case of the
$\nu=1/2 $ system we have for low temperatures  \cite{ho1}
$ \rho^*= m \mu/(2\pi) +O(e^{-\beta \mu }) $. The grand canonical
  potential is then given by (the exact form of $ \Omega $
  in the Hartree-Fock approximation will be derived later)
 \begin{eqnarray}
& & \Omega(\mu,\beta)=  
\Omega_B(\mu,\beta,(2\pi \tilde{\phi} \rho^*))
+ a_1 \, e^2 m^{\frac{3}{2}} \mu^{\frac{3}{2}}
\nts{1} + \nts{1} a_2\, e^2 m^{\frac{3}{2}} \mu^{\frac{1}{2}} \frac{1}{\beta} \nonumber  \\ 
& & +a_3 \, e^2 m^{\frac{3}{2}} \frac{1}{\mu^{\frac{1}{2}}} \frac{1}{\beta^2}
+a_4 \, e^2 m^{\frac{3}{2}} \frac{1}{\mu^{\frac{1}{2}}} 
\frac{1}{\beta^2} \log\left(\frac{1}{\mu \beta}\right) \nonumber \\
& & +O(e^2/(\mu^{3/2} \beta^3))+O(e^4 m^2 \mu^{\frac{1}{2}}) 
.\eqlabel{5560} 
\end{eqnarray}
In the following we will denote
$ \Omega(\mu,\beta)-\Omega_B(\mu,\beta,(2\pi \tilde{\phi} \rho^*)) $
by $ \Omega_{c}(\mu,\beta) $.
We now make the following ansatz for the terms of  $ \mu $
which scales polynomial with 
$ 1/\beta$:  
\begin{eqnarray}
& & \mu (\rho,\beta)  =  
\pi \tilde{\phi} \frac{\rho}{m} +b_1 \,e^2  \rho^{\frac{1}{2}}
+b_2 \, e^2 \frac{m}{ \rho^{\frac{1}{2}}} \frac{1}{\beta} \nonumber \\  
 & & +b_3 e^2 \frac{m^{2}}{\rho^{\frac{3}{2}}} \frac{1}{\beta^2} 
 +b_4 e^2  \frac{m^{2}}{\rho^{\frac{3}{2}}} 
 \frac{1}{\beta^2} \log\left(\frac{m}{\rho \beta}\right) \;.\eqlabel{5570} 
\end{eqnarray}
In the following we will denote the first term in the sum
of $ \mu (\rho,\beta)$ by $ \mu_0\rho,\beta)$.
$ \mu (\rho,\beta)-\mu_0(\rho,\beta) $ will be denoted by $ \mu_{c}(\rho,\beta) $.
We get for $\mu_{c}$ inserting (\ref{1460}), (\ref{1480}),
(\ref{5560}), (\ref{5570}) 
in (\ref{1440}) and taking into consideration
$ \rho^*= m \mu/(2\pi) +O(e^{-\beta \mu}) $:
\begin{eqnarray}
  & & -\frac{\partial \mu_0 }{\partial \rho} 
\frac{\partial \Omega_{c} }{\partial \mu}+\nts{2}\left(\frac{1}{2}\left(   
\frac{\partial (\beta\Omega_{c}) }{\partial \beta}+\mu_{c} \rho
\right)+\Omega_{c}\right) \nts{1}\frac{1}{\rho} \nonumber \\
& &
+ \left(\frac{\mu_{c} }{\rho}-\frac{\partial \mu_{c} }{\partial \rho}
\right) \frac{\Omega_{B}}{\mu}= \rho \frac{\partial \mu_{c} }{\partial \rho}.
\eqlabel{5580}
\end{eqnarray}
We will solve this equation by comparing the coefficients of equal 
powers of $ \beta $. 
If one inspects the equation of the coefficient of $ (\beta)^0 $ in
(\ref{5580})  
one notices that the coefficient $ b_1 $ vanishes in this equation
and one gets a trivial identity which is fulfilled for every chosen $ b_1 $.    
The reason for this indeterminacy of $ b_1 $ is given by the partition of
$ \Omega $ in a sum of a pure Chern-Simons term and $ e^2 $-corrections to
$ \Omega $ (this could be seen by using (\ref{1370}), (\ref{1490})).
Since this partition is not correct we have no equation to determine
the coefficient $ b_1 $. For this reason we have to start properly with a
Chern-Simons perturbation theory which shows a more complicated mean field.
This will be done in a later publication.
We now set $ b_1=0 $. This could be justified by the Luttinger-Ward theorem
\cite{lu1} as a good approximation calculating the $ T=0 $ energy  
$ \Omega $ in RPA because the
RPA does not contain any anomalous graphs \cite{di1,di2}.
By comparing the coefficients of the 
higher order powers of $ 1/\beta$ in (\ref{5580}) 
we get for the $ \nu=1/2 $ system  
\begin{eqnarray}
& & \mu= 2 \pi \frac{\rho}{m}
  -a_2 e^2  \frac{\sqrt{2 \pi}}{2}   
\frac{m}{\rho^{\frac{1}{2}}} \frac{1}{\beta}    
-\left(\frac{a_3}{2} +\frac{3 a_2 \log(2) }{4}  \right) \nonumber \\
& & \times \; e^2 \frac{1}{\sqrt{2 \pi}}\frac{m^2}{\rho^{\frac{3}{2}}}
\frac{1}{\beta^2}
-a_4 e^2  \frac{1}{2 \sqrt{2 \pi}}\frac{ m^2}{\rho^{\frac{3}{2}}}
\frac{1}{\beta^2}\log\left(\frac{m}{2 \pi \rho \beta}\right) \,.
\eqlabel {5610}  
\end{eqnarray}
With the help of the equations (\ref{1450}), (\ref{1460}) and (\ref{1480})
the energy density up to order $e^2$ is given by  
\begin{eqnarray}
U & = & 2 \pi \frac{\rho^2}{m}+ \frac{\partial (\beta \Omega_{c}) }{\partial \beta}
-\beta 2 \pi \frac{\rho}{m}
\left.\frac{\partial\rho_{c}(\mu,\beta)}{\partial\beta}
  \right|_{\mu=\mu(\rho,\beta)}
\eqlabel{5600} \\
& &  - \beta \,\frac{\Omega_{B}}{\rho} \left.\frac{\partial\rho_{c}(\mu,\beta)}{\partial\beta}\right|_{\mu=\mu(\rho,\beta)}  +\mu_{c}
\rho \;. \nonumber 
\end{eqnarray}
$ \rho_{c}(\mu,\beta) $ in (\ref{5600}) is defined by  
$ \rho(\mu,\beta)-m \mu/(2 \pi)$.  
With the help of  (\ref{5560}), (\ref{5600}) we get for the energy of the
$ \nu=1/2 $ system 
\begin{eqnarray}
& &  U  =  (2 \pi) \frac{\rho^2}{m}+a_1 e^2 
(2\pi \rho)^\frac{3}{2}
-\left(\frac{(a_3+a_4)}{2}+\frac{a_2 \log(2)}{4} \right)  \nonumber \\
& & \times 
e^2 \frac{m^2}{(2 \pi \rho)^{\frac{1}{2}}}\frac{1}{\beta^2}
-a_4 e^2 \frac{m^2}{2 (2 \pi \rho)^{\frac{1}{2}}} 
\log\left(\frac{m}{2 \pi \rho \beta}\right)\frac{1}{\beta^2}
\;.\eqlabel{5620}
\end{eqnarray}
In the following we will calculate the Coulomb part of
the Hartree-Fock approximation of
$ H_{CS}$ (\ref{1050}). This term is equal to the Coulomb exchange
Feynman graph given by  
 \begin{equation} \eqlabel{5140}
 \mbox{Ex}^{ee}=\frac{1}{2\,(2\pi)^3}\int d^2k d^2k' \,n_F(k)\,
 n_F(k')\,V^{ee}(|\vec{k}-\vec{k}'|).
 \end{equation}
$ n_F(k) $  is the fermi factor $n_F(k)=1/(1+e^{\beta(k^2/(2m)-\mu)}) $.
$ \mbox{Ex}^{ee} $ is also the $ e^2 $-part of the grand canonical potential 
of the (two dimensional) Coulomb gas. This term was calculated by Isihara and
Toyoda \cite{isi3} within the calculation of the grand canonical potential of the
two dimensional electron gas. Up to order $ 1/\beta^2 $ it is given
by
\begin{eqnarray}
 \mbox{Ex}^{ee} &\approx &  -0.095 \,e^2 m^{\frac{3}{2}}\mu^{\frac{3}{2}}
 \eqlabel{5540}   \\
& &  + \bigg(0.035 -0.0588\log\left(\frac{1}{\mu \beta}\right) \bigg) e^2
\frac{m^{\frac{3}{2}}}{\mu^{\frac{1}{2}}} \frac{1}{\beta^2} 
 \;.  \nonumber
\end{eqnarray} 
By using (\ref{5560}), (\ref{5610}), (\ref{5620}) and (\ref{5540}) we get
for the chemical potential $ \mu^{\scriptsize \mbox{HF}} $
and the energy  $ U^{\scriptsize \mbox{HF}} $ in the Hartree-Fock approximation
\begin{eqnarray}
\mu^{\scriptsize \mbox{HF}} & = & 2 \pi \frac{\rho}{m}
-0.017 \, e^2  \frac{1}{\sqrt{2 \pi}}
\frac{m^2}{\rho^{\frac{3}{2}}}
\frac{1}{\beta^2}  \nonumber   \\
& & +0.029 \, e^2  \frac{1}{\sqrt{2 \pi}}
\frac{m^2}{\rho^{\frac{3}{2}}}
\frac{1}{\beta^2}\log\left(\frac{m}{2 \pi \rho \beta}\right)
,  \eqlabel{5630}  \\
U^{\scriptsize \mbox{HF}} & =  & (2 \pi) \frac{\rho^2}{m}-0.095 \, e^2 
(2\pi \rho)^\frac{3}{2}
+0.012 \, e^2 \frac{m^2}{(2 \pi \rho)^{\frac{1}{2}}}\frac{1}{\beta^2}  \nonumber    \\
& & +0.029 e^2 \frac{m^2}{(2 \pi \rho)^{\frac{1}{2}}}\frac{1}{\beta^2}
\log\left(\frac{m}{2\pi \rho \beta}\right) \;. \eqlabel{5640} 
\end{eqnarray}
For $ T=0 $ the Coulomb part of $U^{\scriptsize \mbox{HF}} $ 
is in good agreement
with the Coulomb energy of numerical simulation of electrons
on a sphere by Morf and d'Ambrumenil \cite{mo1} and by Girlich \cite{gi1}
($ \approx -0.1 \,e^2 (2\pi \rho)^{3/2} $).  
We can now compare the $ e^2 $-part of the energy of the $ \nu=1/2 $ system
with the $ e^2 $-part of the energy of 
the Coulomb system including the spin degree of freedom. 
The energy up to order $ e^2 $ of the two dimensional Coulomb
system was calculated by Isihara and Toyoda \cite{isi3}.  
It is calculated from the grand canonical potential up to order $ e^2$ 
which is the summation of the grand canonical potential of an 
interaction free electron gas and the 
Coulomb exchange graph. 
Isihara and Toyoda calculated for this energy 
\begin{eqnarray}\eqlabel{5650}
& & U_{\mbox{\tiny Coul}} =  \frac{\pi}{2} \frac{\rho^2}{m}+\frac{\pi}{12}m\frac{1}{\beta^2}
-0.067\, e^2 
(2\pi \rho)^\frac{3}{2}\\ 
& &+0.063 \, e^2 \frac{m^2}{(2 \pi \rho)^{\frac{1}{2}}}\frac{1}{\beta^2}
+0.166 \, e^2  \frac{m^2}{(2 \pi \rho)^{\frac{1}{2}}}\frac{1}{\beta^2}
\log\left(\frac{m}{\pi \rho \beta}\right) \;. \nonumber 
\end{eqnarray}   
In figure \ref{energ} we show the $ e^2 $-Coulomb energy 
of the $ \nu=1/2 $ system in Hartree-Fock approximation as well as
the $ e^2 $-Coulomb energy of
the two dimensional Coulomb system (the spin polarized system as well as 
the system including the spin degree of freedom).
It is seen from these curves that the energy curve of the CS gas
has a much less steeper slope in comparison to the Coulomb gas including the 
spin degree of freedom.
Furthermore, one sees from the figure that the main responsibility for this behaviour
is the spin polarization of the CS gas. But also the $ \mu $ and energy
correction formulas of the CS gas of section II increase this effect.
The flatness of the energy curve can be understood by perturbational
arguments. As mentioned below (\ref{1460}) $ \langle H_0(B) \rangle $ does not
contribute to the energy in $ e^2 $-order for $ T \ge 0 $. Thus the energy
is given by the average value $ \langle H_{ee}(B) \rangle $ calculated
with respect to the Gibb's weight times the eigen functions
of $ H_0(B)+ H_{ee}(B) $ which result  
from the lowest Landau level by Coulomb perturbations. 
It is easily seen that  this term does not have any temperature corrections
up to order $ e^2 $.
Therefore, we get that the $ \nu=1/2 $ Chern-Simons system does only pick up 
temperature corrections in higher $ O(e^4) $ order in the parameter range
(\ref{10}). Summarizing, we get an agreement of the perturbational calculated
energy (\ref{5640}) with this exact result. 

\subsection{The chemical potential and the energy
  of the \boldmath $ \nu=1/2 $ \unboldmath Chern-Simons
  system for \boldmath $ 1/\beta \, \ll \,  e^2 \sqrt{\rho}   \,  \ll  \,
 \rho/m $ \unboldmath}

As mentioned in the introduction of this paper the chemical potential and the
energy in the parameter range 
(\ref{15}) can not be calculated perturbationally through the Chern-Simons
theory. HLR \cite{hlr} and Kim, Lee \cite{ki2} used the RPA
for a resummation of diagrams to 
calculate the temperature corrections
of the grand canonical potential in this parameter range.
We will use in the following the result of their  calculation to obtain
the chemical potential and the energy of the $\nu=1/2 $ system.
As in the last subsection we do not determine the first term in the
expansion of the grand canonical potential in the parameter range
(\ref{15}) perturbationally. 
This term is given by the grand canonical potential of an electron gas
in the lowest Landau level. This can be calculated through $
\Omega_B(\mu.\beta,(2 \pi \tilde{\phi} \rho^*)) $ limiting the summation
in (\ref{1510}) to the lowest Landau level ($n=0$). Since the higher Landau
level terms contribute to the chemical potential and the
energy with  exponentially vanishing
temperature corrections we will use in the following the full
$\Omega_B(\mu.\beta,(2 \pi \tilde{\phi} \rho^*)) $.
Thus the grand canonical potential of the $\nu=1/2 $ system in the parameter
range (\ref{15}) is of the following form
\begin{eqnarray}
& & \Omega(\mu,\beta)=  
\Omega_B(\mu,\beta,(2\pi \tilde{\phi} \rho^*))+a'_1 \,
e^2 m^{\frac{3}{2}} \mu^{\frac{3}{2}}
+a'_2\, m  \mu \frac{1}{\beta} \nonumber  \\ 
& & +a'_3 \, \frac{1}{e^2}  m^{\frac{1}{2}} \mu^{\frac{1}{2}} \frac{1}{\beta^2}
+a'_4 \, \frac{1}{e^2} m^{\frac{1}{2}} \mu^{\frac{1}{2}}
\frac{1}{\beta^2} \log\left(\frac{1}{e^2 \sqrt{m \mu} \beta}\right) 
\nonumber \\
& & +
O(1/ (e^4 \beta^3))+O(e^4 m^2 \mu) 
.\eqlabel{5670} 
\end{eqnarray}
We now make the following ansatz for the terms of  $ \mu $
which scales polynomial with $ 1/\beta$:  
\begin{eqnarray}
& & \mu (\rho,\beta)  =  
\pi \tilde{\phi} \frac{\rho}{m} +b'_1 \,e^2  \rho^{\frac{1}{2}}
+b'_2 \, \frac{1}{\beta} \nonumber \\  
 & & +b'_3 \, \frac{1}{e^2} \frac{1}{\rho^{\frac{1}{2}}} \frac{1}{\beta^2} 
 +b'_4 \, \frac{1}{e^2} \frac{1}{\rho^{\frac{1}{2}}} \frac{1}{\beta^2} 
 \log\left(\frac{1}{e^2 \sqrt{\rho} \beta}\right) \;.\eqlabel{5680} 
\end{eqnarray}
Using the results of the last section (see the discussion below (\ref{1480}))
it is easy to see
that the equations (\ref{5580}) and (\ref{5600}), determining the
chemical  potential and the
energy are still valid for the approximation (\ref{5670}), (\ref{5680}).
By using these equations we get with the help of the grand canonical potential
(\ref{5670}) for the $ \nu=1/2 $ system
(as in the last subsection we set $ b'_1=0 $) 
\begin{equation}
 \mu= 2 \pi \frac{\rho}{m}\,.
\eqlabel {5690}  
\end{equation}
The energy of the $ \nu=1/2 $ system is given by
\begin{eqnarray}
& &  U  =  (2 \pi) \frac{\rho^2}{m}+a'_1 e^2 
(2\pi \rho)^\frac{3}{2}
-\left(a'_3+ a'_4\right) \frac{1}{e^2}
(2 \pi \rho)^{\frac{1}{2}} \frac{1}{\beta^2}   \nonumber \\
& &- a'_4  \frac{1}{e^2}
(2 \pi \rho)^{\frac{1}{2}} \frac{1}{\beta^2}
\log\left(\frac{1}{e^2 \sqrt{\rho} \beta}\right)
\;.\eqlabel{5700}
\end{eqnarray}

As mentioned above HLR \cite{hlr} and Kim, Lee \cite{ki2} calculated the
temperature correction
terms in (\ref{5670}) through a resummation of the RPA. The $ e^2 $-order term
of the grand canonical potential in RPA for temperature
$ T=0 $ was calculated by us in \cite{di1}. By using these
two results we get for the grand canonical potential (\ref{5670}) 
 \begin{eqnarray}
& & \Omega_{\scriptsize \mbox{RPA}}(\mu,\beta)=
\Omega_B(\mu,\beta,(2\pi \tilde{\phi} \rho^*))-0.13 \, e^2
 m^{\frac{3}{2}} \mu^{\frac{3}{2}} \eqlabel{5660}  \\
& & - \frac{1}{6} \frac{1}{e^2}
(2m\mu)^{\frac{1}{2}}
\frac{1}{\beta^2}
\log\left(\frac{1}{ e^2 \sqrt{m \mu} \beta}\right) 
+ O\left(\frac{1}{e^2 \sqrt{m \mu} \beta^2}\right)\,. \nonumber 
\end{eqnarray}
By using this  grand canonical potential we obtain from 
(\ref{5690}) and (\ref{5700}) up to the order  $ O((1/\beta^2)
\log(1/e^2 \sqrt{\rho} \beta)) $
\begin{eqnarray}
 \mu^{\scriptsize \mbox{RPA}} & = & 2 \pi  \frac{\rho}{m}
 \,, \eqlabel{5710}  \\
U^{\scriptsize \mbox{RPA}} & = &  2 \pi \frac{\rho^2}{m}
-0.13 \, e^2 \, 
 (2 \pi \rho)^{\frac{3}{2}} \nonumber \\
& &  +\frac{\sqrt{\pi}}{3} \frac{1}{e^2}
\rho^{\frac{1}{2}} \frac{1}{\beta^2}
\log\left(\frac{1}{e^2 \sqrt{\rho} \beta}\right)
\;.              \eqlabel{5720}
\end{eqnarray}
In \cite{ki2} the temperature corrections to the 
ground state energy was calculated
by $ \partial (\beta 
\Omega_{\scriptsize \mbox{RPA}})/ (\partial \beta) \; (2\pi \rho/m,\beta)
$. When comparing this expression 
with the temperature corrections of
$ U^{\scriptsize \mbox{RPA}}$ (\ref{5720}) we get the same result.
This is also correct for the full $ \Omega $ (\ref{5670}).
This is the reason for getting the same effective mass either by
determining the effective mass through the one particle Green's function or
through a comparison  of the specific heat of the $ \nu=1/2 $ system with the
specific heat of an interaction free electron system (one can easily show
from the equations above that this does not depend on the setting $ b_1'=0 $). 
The accordance of these two masses is well known in the case of the Coulomb
system.

\section{Conclusion}

We considered in this paper at first the question whether  corrections
to the well known equations
for determining in a Chern-Simons theory the chemical potential and the energy
from a grand canonical potential in which one has
eliminated the magnetic field and the field of the positive background
through a mean field formalism
should be  taken into
account. 
We showed that one gets 
corrections to these  well known 
equations in contrast to the Coulomb system. We stated explicitly
these corrections. Furthermore, we showed that these corrections
can be determined from    
the grand canonical potential calculated by this mean field formalism.
We should mention that our equations determining 
the chemical potential and the energy from this field eliminated
grand canonical potential
of the $ \nu=1/2 $ system are also valid for
other theories than the Chern-Simons theory of HLR.

With the help of these corrections 
and the well known
result \cite{isi3} for the grand canonical potential of the Coulomb exchange
graph we calculated next the Hartree-Fock energy of the spin polarized 
$ \nu=1/2 $ Chern-Simons system for low temperatures $ T>0 $ up to order $ T^2
$. The parameter range of the calculated  chemical potential
and the energy is given by
$   e^2 \sqrt{\rho} \, \ll  \,  1/\beta \,    \ll  \, \rho/m $.
We compared
the energy with the energy of the two dimensional Coulomb
gas taking into account the spin degrees of freedom. We get that
the spin polarization as well as
the corrections to the
equations determining the 
chemical potential and the energy 
cause the energy as a function of the temperature to flatten.
We showed that the exact behaviour of the energy within our approximation
would have
temperature corrections which are zero.
Next we calculated the chemical potential and the energy from the
temperature corrections of the grand
canonical potential obtained in the RPA \cite{hlr,ki2}.
The parameter range of the chemical potential
and the energy is given by
$ 1/\beta \,\ll \, e^2 \sqrt{\rho}   \, \ll  \, \rho/m $.
We showed that one gets the same result for the energy as in the
simplification \cite{ki2} by using the chemical potential 
of the interaction free $ \nu=1/2 $ system for temperature $ T=0 $
and neglecting the corrections
in the equation for determining the energy from the grand canonical
potential. 

Finally we mention that it should be in
principle possible to calculate the energy of the $ \nu=1/2 $  system for
temperatures $ T>0 $ through temperature heat capacity measurements.
This was earlier done by Bayot et al. for the $ \nu \approx 1 $ system
\cite{ba1}.

\bigskip
We would like to thank K.~Luig, W.~Apel and W.~Weller for many helpful 
discussions during the course of this work. 
Further we have to acknowledge the financial support by the Deutsche
Forschungsgemeinschaft, Graduiertenkolleg "Quantenfeldtheorie".

\end{multicols}

\end{document}